\documentclass[usegraphicx]{mn2e}
\usepackage{epsf}
\usepackage{color}

\newcommand{\msun}{\,{\rm M_\odot}}
\newcommand{\rhob}{\rho_{\rm b}}
\newcommand{\etal}{{et al.\ }}
\newcommand{\ie}{{i.e.\ }}
\newcommand{\eg}{{e.g.,\ }}
\newcommand{\Om}{\Omega}
\newcommand{\hinv}{h^{-1}}
\newcommand{\de}{\delta}

\newcommand{\lcdm}{$\Lambda$CDM~}

\newcommand{\Mvir}{M_{\rm vir}}
\newcommand{\Rvir}{R_{\rm vir}}
\newcommand{\cvir}{c_{\rm vir}}
\newcommand{\wQ}{w}
\newcommand{\deltac}{\delta_{\rm c}}
\newcommand{\DeltaVir}{\Delta_{\rm vir}}
\newcommand{\sigc}{\sigma_{\log c}}
\newcommand{\Lbox}{L_{\rm box}}

\newcommand{\be}{\begin{equation}}
\newcommand{\ee}{\end{equation}}
\newcommand{\ba}{\begin{eqnarray}}
\newcommand{\ea}{\end{eqnarray}}
\newcommand{\brr}{\begin{array}}
\newcommand{\err}{\end{array}}
\newcommand{\bc}{\begin{center}}
\newcommand{\ec}{\end{center}}
\newcommand{\f}{\frac}

\newcommand{\VQQ}{V,_{\phi \phi}}

\title{Dark Energy and Dark Matter Haloes}
\author[
Kuhlen et al.]{Michael Kuhlen$^1$, Louis E. Strigari$^2$, 
Andrew R. Zentner$^{3,4}$, James S. Bullock$^5$\thanks{Hubble Fellow}
\newauthor
and  Joel R. Primack$^6$\\
$^1$Department of Astronomy and Astrophysics, University of California at Santa Cruz, 1156 High Street, Santa Cruz, CA 95064 \\
$^2$Department of Physics, The Ohio State University, 174 W. 18th Avenue, Columbus, OH 43210\\
$^3$Center for Cosmological Physics, The University of Chicago, 5640 S. Ellis Avenue, Chicago, IL 60637\\
$^4$Department of Astronomy and Astrophysics, The University of Chicago, 5640 S. Ellis Avenue, Chicago, IL 60637\\ 
$^5$Harvard-Smithsonian Center for Astrophysics, 60 Garden Street, Cambridge, MA 02138\\
$^6$Physics Department, University of California at Santa Cruz, 1156 High Street, CA 95064}
\date{Submitted 2004 April 05}

\pagerange{\pageref{firstpage}--\pageref{lastpage}} \pubyear{2004}

\begin{document}

\label{firstpage}

\maketitle

\begin{abstract}
We investigate the effect of dark energy on the density profiles of
dark matter haloes with a suite of cosmological N-body simulations and
use our results to test analytic models.  We consider constant
equation of state models, and allow both $\wQ \ge -1$ and $\wQ<-1$.
Using five simulations with $\wQ$ ranging from $-1.5$ to $-0.5$, and
with more than $\sim 1600$ well-resolved haloes each, we show that the
halo concentration model of Bullock \etal (2001) accurately predicts
the median concentrations of haloes over the range of $\wQ$, halo
masses, and redshifts that we are capable of probing.  We find that
the Bullock \etal (2001) model works best when halo masses and
concentrations are defined relative to an outer radius set by a
cosmology-dependent virial overdensity.  For a fixed power spectrum
normalization and fixed-mass haloes, larger values of $\wQ$ lead to
higher concentrations and higher halo central densities, both because
collapse occurs earlier and because haloes have higher virial
densities.  While precise predictions of halo densities are quite
sensitive to various uncertainties, we make broad comparisons to
galaxy rotation curve data.  At fixed power spectrum normalization
(fixed $\sigma_8$), $\wQ > -1$ quintessence models seem to exacerbate
the central density problem relative to the standard $\wQ= -1$
model. For example, models with $w \simeq -0.5$ seem disfavored by the
data, which can be matched only by allowing extremely low
normalizations, $\sigma_8 \la 0.6$.  Meanwhile $\wQ < -1$ models help
to reduce the apparent discrepancy. We confirm that the Jenkins \etal
(2001) halo mass function provides an excellent approximation to the
abundance of haloes in our simulations and extend its region of
validity to include models with $\wQ < -1$.
\end{abstract}


\section{Introduction}

In the prevailing model of galaxy formation, galaxies assemble and
evolve in the potential wells established by gravitationally bound
haloes of cold and collisionless dark matter (CDM).  Except for some
possible difficulties on small scales, the CDM model is remarkably
successful in explaining a large number of observations.  However,
this success requires an additional ``dark energy'' component, that
drives an accelerated cosmic expansion, to be added to the universal
energy budget.  While the presence of dark energy is firmly
established observationally, measuring its equation of state as well
as developing a theoretical understanding of the nature of the dark
energy are two of the biggest outstanding problems in cosmology today.
Dark energy not only affects the large-scale evolution of the
Universe, but also the collapse histories and density structures of
dark matter haloes.  Understanding the precise nature of these effects
is important for studies that aim to quantify the nature of dark
energy using strong (e.g., Sarbu, Rusin, \& Ma 2001; Huterer \& Ma
2004; Kuhlen, Keeton, \& Madau 2004) and weak (e.g.  Hu \& Jain 2003;
Bartelmann et al. 2002) gravitational lensing.  Changing the dark
energy model should similarly change expectations for galaxy rotation
curves, and could affect one of the main small-scale problems facing
CDM -- the central density problem (\eg Zentner \& Bullock 2002;
McGaugh, Barker, \& de Blok 2003, and references therein).  In the
present paper, we use a suite of N-body simulations to study how halo
density profiles change as a function of dark energy equation of
state, discuss our results in the context of analytic models, and
discuss the observational implications of dark energy on galaxy
scales.

The existence of some form of dark energy is supported by a
preponderance of data.  Taken together, observations of the
magnitude-redshift relation of type Ia supernovae (SNIa; Perlmutter
\etal 1999; Riess \etal 2001; Knop \etal 2003; Barris \etal 2003), the
power spectrum of cosmic microwave background (CMB) anisotropy
(Spergel \etal 2003; Tegmark \etal 2003a), the power spectrum of
galaxy clustering (Dodelson \etal 2003; Tegmark \etal 2003b), and the
luminosity function and baryon fraction of clusters (Allen \etal 2003)
provide nearly unimpeachable evidence for the existence of dark
energy.  The most common supposition is that the dark energy takes the
form of a cosmological constant or vacuum energy.  In this case, the
energy density $\rho$, and pressure $p$, are related through $p =
-\rho$.  An attractive alternative candidate for the dark energy is
the potential energy of a slowly-varying scalar field $\phi$, or
``quintessence'' (\eg Ratra \& Peebles 1988; Caldwell, Dav{\'e}, \&
Steinhardt 1998).

A convenient parametrization of the dark energy is through an equation
of state $\wQ \equiv p_{\phi}/\rho_{\phi}$ relating its energy density
and pressure.  In general, the equation of state parameter $\wQ$ is
time-varying, but it is useful to model quintessence with a constant
equation of state parameter because current observational data sets
have limited power to distinguish between a time-varying and constant
equation of state (\eg Kujat et al. 2002).  Useful limits on the
equation of state for the dark energy, assuming that it remains
constant in time, come from SNIa studies, $-1.67 < \wQ < -0.62$
($2\sigma$; Knop \etal 2003), and can be refined by combining SNIa
data with CMB anisotropy and galaxy clustering statistics yielding
$-1.33 < \wQ < -0.79$ at $2\sigma$ (Tegmark \etal 2003b). For our
simulations, we adopt an empirical view and study five models with
constant $\wQ$, that span a comparably large range of parameter space:
$\wQ = -1.5, -1.25, -1.0, -0.75, -0.5$.

Our theoretical understanding of halo profiles has improved recently
largely through numerical simulations, performed in the context of CDM
plus cosmological constant ($\Lambda$CDM) or standard CDM (SCDM, \ie
$\Omega_{\rm M} = 1$, $\Omega_{\rm Q}=0$) cosmologies (Navarro, Frenk,
\& White, 1995, 1996, 1997, hereafter NFW; Kravtsov, Klypin, \&
Khokhlov 1997; Ghigna \etal 1998; Jing 2000; Bullock \etal 2001; Eke,
Navarro, \& Steinmetz 2001; Wechsler et al. 2002, hereafter W02; Zhao
et al. 2003; Hayashi \etal 2003; Navarro \etal 2003; for a review, see
Primack 2003).  It is generally understood that the final density
profiles of haloes are linked closely to their formation histories.
Halo central densities are set during an early, rapid-accretion phase,
and tend to be proportional to the density of haloes in the Universe
at the time of this rapid collapse (W02; Zhao et al. 2003; Tasitsiomi
et al. 2003).  Larger values of $\wQ$ lead to earlier collapse times
and also to more rapid collapse of overdensities, thus we expect
haloes with higher relative central densities and higher virial
densities (see our discussions in \S~\ref{sec:Qstruc} and
\S~\ref{sec:halostruc}).  In \S\S~\ref{sec:sims}-\ref{sec:DeltaVby2}
we quantify these effects and test the expected scalings.  We verify
that the analytic technique of Bullock \etal (2001, B01) for
predicting halo concentrations works well when applied directly to
constant $\wQ$ cosmologies, implying that it is fairly generally
applicable.

A related issue is the effect of dark energy on the halo mass
function.  Although it is not directly observable, theoretical insight
into the expected halo mass function is attainable through current
N-body simulations.  The effects of dark energy on halo mass functions
have been investigated by Bartelmann, Perota, \& Baccigalupi (2003),
Linder \& Jenkins (2003), Klypin \etal (2003), Macci\`o \etal (2003),
and \L okas, Bode, \& Hoffman (2003) and the accurate prediction of
halo mass functions as well as accretion histories and density
structures over a large range of redshifts is necessary in order to
take full advantage of the ability of upcoming cluster surveys, such
as the Sunyaev-Zeldovich Array (\texttt{\small
http://astro.uchicago.edu/sza/}), to constrain the dark energy
equation of state (see Carlstrom, Holder, \& Reese 2001 for a review).
There appears to be general agreement that halo mass functions can be
approximated accurately by the ``universal'' mass function of Jenkins
\etal (2001, hereafter J01) even in models with dark energy.  In
\S~\ref{sec:mf}, we confirm the results of previous studies of
quintessence cosmologies with $\wQ > -1$, and extend this agreement to
models with $\wQ < -1$.

Klypin \etal (2003) and Dolag \etal (2003) have performed previous
numerical studies of CDM haloes in quintessence cosmologies.  Where
our study overlaps with these, our results are generally in agreement.
This study extends, complements, and improves upon previous studies in
several ways.  We have investigated the effects of normalizing the
power spectrum using clusters vs. CMB, which yield quite different
results. We have assembled large catalogues of haloes, with masses and
concentrations for more than $1600$ haloes with more than $100$
particles each in each simulation. With this large number of haloes we
are able to measure the distribution of concentrations at fixed mass
and test the predictions of the B01 model for both the mean relation
between concentration and mass and the halo-to-halo scatter.  We also
extend the range of quintessence parameter space probed by simulations
by exploring two models with $\wQ<-1$.

The format of this paper is as follows.  After a brief review of the
basics of structure formation in models with dark energy in
\S~\ref{sec:Qstruc}, we go on in \S~\ref{sec:halostruc} to describe
modifications to the B01 model for halo concentrations in quintessence
cosmologies.  In \S~\ref{sec:sims}, we describe the set of numerical
simulations that we performed.  In \S~\ref{sec:nbody_results}, we
present the mass functions and concentrations of CDM haloes in these
simulations and compare them to the results of the analytic B01 model.
In \S~\ref{sec:DeltaVby2}, we briefly discuss the central density
problem of CDM in light of our results.  In \S~\ref{sec:conclusions},
we present a summary of our findings, highlight improvements and
differences to previous studies, and discuss the implications of this
work.  Throughout this paper, we assume a flat universe with present
matter density relative to critical of $\Omega_{\rm M}=0.3$, and
Hubble parameter $h=0.7$. The quintessence energy density is then
$\Omega_{\rm Q} = 1 - \Omega_{\rm M} = 0.7$.  We refer to a model with
a cosmological constant ($\wQ=-1$) as $\Lambda$CDM. We use the terms
``quintessence'' and ``$Q$CDM'' to refer to all models which have $\wQ
\ne -1$.

\section{Structure formation in Quintessence Cosmologies}
\label{sec:Qstruc}

In this Section, we present a brief overview of the growth of
structure in quintessence cosmologies.  In \S\ref{sec:pert}, we
introduce the basic results of cosmological perturbation theory with
quintessence and discuss our computations of the matter power
spectrum.  We briefly discuss the normalization of the power spectrum
in \S \ref{sec:norm}.  A convenient way to define the mass and radius
of a dark matter halo is through a mean overdensity $\Delta_{\rm
vir}$, relative to the background density (see
\S\ref{sec:B01model}). The idea is to choose $\Delta_{\rm vir}$ so as
to delineate the boundary between virialized and in-falling material.
The equivalent linear overdensity at collapse $\delta_{\rm c}$, is a
quantity that is used to delineate the mass scale of objects that are
forming at a particular redshift.  Often, these overdensity criteria
are chosen with reference to the evolution of a spherical tophat
overdensity.  In \S\ref{sec:tophat}, we discuss the spherical tophat
model in quintessence cosmologies.

We explore both conventional quintessence models with $\wQ \ge -1$
and, adopting an empirical approach, we pursue models with $\wQ < -1$.
Requiring the kinetic energy of the quintessence to be positive
imposes the condition $\wQ \ge -1$ on the quintessence equation of
state. When $\wQ \ge -1$ and constant, all computations can be
performed knowing only the value of $\wQ$ by assuming that the scalar
field Lagrangian has a canonical kinetic term.  To explore the
parameter space $\wQ < -1$ self-consistently, we must choose a
particular model for scalar field dynamics and one simple possibility
is a Lagrangian with a non-canonical kinetic term that differs from
the canonical case by a negative sign.  Explicitly, we adopt the
``phantom energy'' Lagrangian for the field, $\phi$: $\cal{L} = -
\partial_{\mu} \phi \partial^{\mu} \phi - {\rm V}(\phi)$ (Caldwell
2002; Carroll, Hoffmann, \& Trodden 2003; Cline, Jeon, \& Moore 2003).
Having made this choice, we can express the derivatives of the
potential completely in terms of $\wQ$ (Dave, Caldwell, \& Steinhardt
2002).

\subsection{Cosmological Perturbations} 
\label{sec:pert} 

Here we consider linear perturbations to the CDM density field,
$\delta(\vec{x}) \equiv \delta \rho(\vec{x})/ \rhob$, with $\rhob$ the
mean density of dark matter in the universe.  The linear evolution of
$\delta(\vec{x})$ follows from solving the linearized
Einstein-Boltzmann equations (\eg Ma \& Bertschinger 1995).  We
perform our calculations in the synchronous gauge with a modified
version of the publicly-available Einstein-Boltzmann code CMBfast by
Seljak \& Zaldarriaga (1996).  We assume a scale-invariant spectrum of
adiabatic primordial density fluctuations and a baryon density
$\Omega_{\rm B}h^{2}=0.02$.

\begin{figure}
\includegraphics[width=84mm]{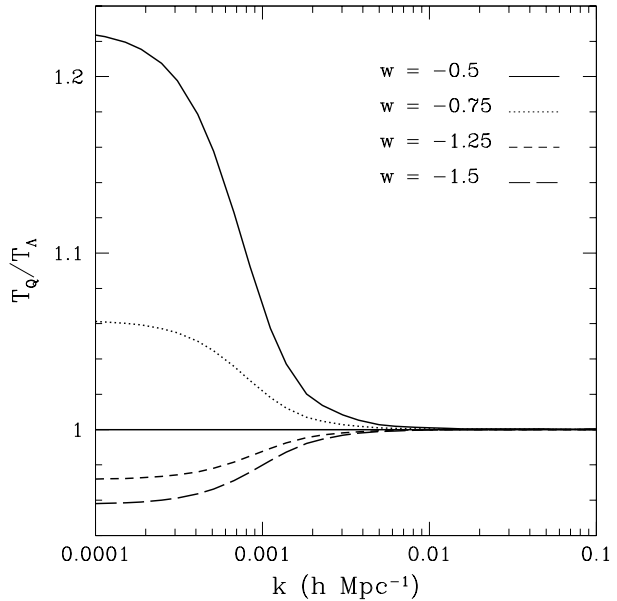}
\caption{
The ratio of $z=0$ quintessence transfer functions to \lcdm for
different models. The transfer functions are similar for wavenumbers
$k \ga 0.01 \, h \, {\rm Mpc}^{-1}$.  The different equation of state
parameters are shown in the legend.
}
\label{fig:tfratio}
\end{figure}

The linear equation for fluctuations in the quintessence field in
Fourier space is

\be
\label{eq:Qlinear} 
\ddot{\delta \phi} + 3 H \dot{\delta \phi} + ( k^{2}/a^2 \pm \VQQ) \delta \phi
= \dot{\delta_{k}}[ ( 1 + w_{Q}) \rho_{\phi} ]^{1/2},  
\ee

\noindent where $\VQQ \equiv \partial^{2} V/\partial \phi^{2}$,
$\rho_{\phi}$ is the mean energy density in the quintessence field,
and the plus (minus) signs correspond to fields with positive
(negative) kinetic energy. Here the dots represent derivatives with
respect to cosmological proper time $t$, and $\delta_{k}$ is the
linear CDM fluctuation in Fourier space. The power spectrum of linear
density perturbations is defined as $P(k) \equiv \langle |\delta_{k}|
^{2} \rangle$. The CDM transfer functions for our scale-invariant
primordial spectrum are defined from

\be
\label{eq:tf}
P(k,z) = A_{Q} k T^{2} (k,z) \frac{D^{2} (z)}{D^{2}(0)},  
\ee

\noindent where $A_{Q}$ is the normalization and $D(z) \equiv \delta
(z) / \delta (0)$ is the linear growth factor.  On scales larger than
the Compton wavenumber, $k_Q \sim \sqrt{\VQQ}$, $\delta \phi$ can grow
and source the evolution of $\delta_{k}$. Contrarily, on small scales
$k \, \ga \, k_Q$, perturbations in the quintessence field decay, so
the $Q$CDM transfer functions have the same form as those of standard
$\Lambda$CDM, while on scales $k \, \la \, k_Q$, the transfer
functions reflect the clustering of $\phi$ (Ma, Caldwell, Bode, \&
Wang 1999).  Figure~\ref{fig:tfratio} shows the ratio of the CDM
transfer functions with quintessence, $T_{Q}$, to the transfer
function in the corresponding $\Lambda$CDM model ($\wQ = -1$),
$T_{\Lambda}$, at $z=0$. For $k \, \la \, k_Q$ and $ w \ge -1$,
perturbations in the quintessence field, $\delta \rho_{\phi} - 3
\delta p_{\phi}$, source the growth of $\delta_{k}$. For $w \le -1$,
the source term from the quintessence field changes sign, resulting in
a relative decrease in $\delta_{k}$.

\subsection{Power Spectrum Normalization}  
\label{sec:norm}

One of the most important parameters determining the scale radii of
halo density profiles is the normalization of the matter power
spectrum on small scales.  In our N-body experiments, we choose to
normalize each $\wQ$ model with similar values of $\sigma_8$, set by
cluster abundance estimates.  We do so because the abundance of
clusters is a more direct probe of the amount of power on the scales
that are relevant to galaxy formation than the CMB anisotropy
measurements and in order to isolate the differences that arise
because of variations in the expansion rate in models with different
values of $\wQ$.
 
We normalize the power spectrum to the abundance of massive clusters
using x-ray flux and temperature measurements from the cores of
clusters of galaxies, and the corresponding conversion to cluster mass
from the mass-temperature ($M$-$T$) relation.  With all other
cosmological parameters fixed, a comparison of the observed mass
function to the predicted mass functions from N-body simulations (J01)
determines the normalization parameter $\sigma_8$.  The quintessence
field is smooth on scales much smaller than the horizon size, thus the
values of $\sigma_8$ derived from $z=0$ cluster measurements are
rather insensitive to $\wQ$.

The largest source of error in the determination of $\sigma_8$ from
clusters is the uncertainty in the normalization of the $M$-$T$
relation.  Simulations consistently predict an $M$-$T$ relation that
is a factor of $\sim 2$ greater than the one observed from x-ray
temperature data (\eg Seljak 2002 and references therein).  For \lcdm
and $\Om_{\rm M} =0.3$, our current understanding of the $M$-$T$
relation places $\sigma_8$ very broadly in the range $0.65 \la
\sigma_8 \la 1.1$ (Pierpaoli, Scott, \& White 2001; Reiprich \&
B\"ohringer 2002; Pierpaoli \etal 2003).  Using the HIFLUGCS cluster
mass function (Reiprich \& B\"ohringer 2002), we find for $\Om_M =
0.3$, $\sigma_8 \simeq 0.74$, nearly independently of $\wQ$ (Kuhlen,
Keeton, \& Madau 2004). We normalize the power spectra to $\sigma_8 =
0.742, 0.740, 0.738, 0.736, 0.734$ for $\wQ = -1.5, -1.25,-1.0, -0.75,
-0.50$ respectively.  In principle, the value of $\sigma_8$ as
determined from clusters is degenerate with $\Om_{\rm M}$
(e.g. Schuecker et al. 2003), indeed the global best fit to the
HILFUGUS sample is $\Om_{\rm M} \simeq 0.12$, $\sigma_8 \simeq 0.96$
in a \lcdm cosmology; however, we adopt values of $\sigma_8$ that best
fit the data given our choice of $\Omega_{\rm M}=0.3$.

\begin{figure}
\includegraphics[width=84mm]{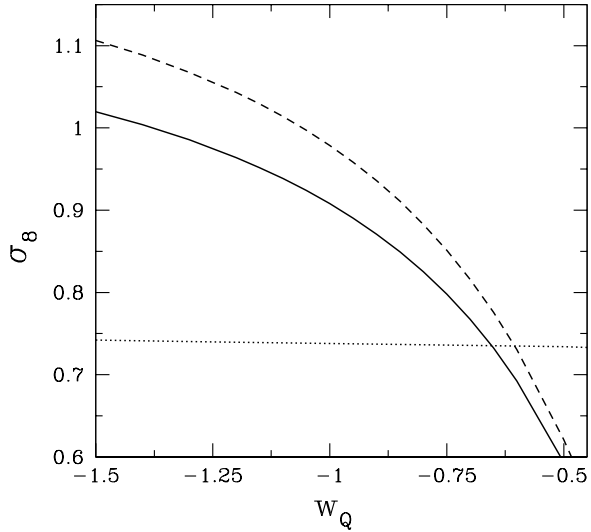}
\caption{
The power spectrum normalization $\sigma_8$, implied by CMB anisotropy
as a function of $\wQ$.  The solid line shows the values of $\sigma_8$
that we obtain by assuming the optical depth to the last scattering
surface $\tau = 0$.  The dashed line shows the values of $\sigma_8$
implied by adopting $\tau = 0.17$.  The dotted line shows the values
of $\sigma_8$ that we infer from the abundance of massive x-ray
clusters.
}
\label{fig:sig8vswQ}
\end{figure}

While we have chosen to normalize our numerical simulations using the
cluster abundance, this normalization is fairly uncertain.  In the
interest of completeness, we remark that not all of these model
normalizations are consistent with $n=1$ normalization to CMB
anisotropy (even modulo uncertainties in the reionization epoch).  In
the simplest case, with all other cosmological parameters held fixed,
the value of $\wQ$ affects the CMB-derived $\sigma_8$ normalization
primarily through the late-time integrated Sachs-Wolfe (ISW) effect
(see Hu \& Sugyiyama 1995).  For universes with $\wQ \ge -1$, the
quintessence energy density becomes comparable to that of CDM at
earlier epochs relative to $\Lambda$CDM, resulting in greater
variation in the gravitational potentials along lines of sight to the
surface of last scattering.  For lower values of $\wQ$, the ISW effect
is not as prominent because quintessence becomes dynamically important
only at more and more recent epochs.  In Figure \ref{fig:sig8vswQ}, we
show the value of $\sigma_8$ implied by the CMB normalization as a
function of the equation of state parameter $\wQ$. As we stated above,
the decrease in $\sigma_8$ is due to the increased importance of the
ISW effect as $\wQ$ increases.

Care must be taken when setting the CMB normalization.  First, we note
that when we performed our N-body experiments we fixed the parameter
$\Omega_{\rm M}=0.3$ for all models rather than allowing $\Omega_{\rm
M}$ to vary along the $\Omega_{\rm M}$-$\wQ$ degeneracy in the angular
diameter distance.  Our most extreme models are currently disfavored
by the present observational data (\eg Tegmark \etal 2003; Knop \etal
2003), but as our intent is to address the effect of quintessence on
structure and halo formation, we do not consider this to be a serious
deficiency.  In principle, the WMAP (Spergel \etal 2003) result of
high optical depth to the last scattering surface $\tau$, has made the
determination of $\sigma_{8}$ from the CMB less robust, as the
scattering off of free electrons damps anisotropies on scales that are
sub-horizon at the epoch of reionization, thereby introducing a
degeneracy between $\tau$ and $\sigma_{8}$.  The dashed line in
Figure~\ref{fig:sig8vswQ} shows the CMB-normalized $\sigma_8$ as a
function of $\wQ$ implied by adopting an optical depth to the last
scattering surface of $\tau = 0.17$, in line with the WMAP
expectations.\footnote{ Note that a high optical depth to reionization
appears to be difficult to reconcile with low values of $\sigma_8 \la
0.75$ (\eg Somerville, Bullock, \& Livio 2003), though the CMB-derived
$\tau$ is very uncertain and is somewhat degenerate with the tilt of
the power spectrum, so tilted models with low $\sigma_8$ may still be
viable.  This issue will likely be settled with future CMB data and
analysis.}  We return to a discussion of the relative importance of
$\sigma_8$ and $\wQ$ in
\S\S~\ref{sec:DeltaVby2}-\ref{sec:conclusions}.
   
\begin{figure}
\includegraphics[width=84mm]{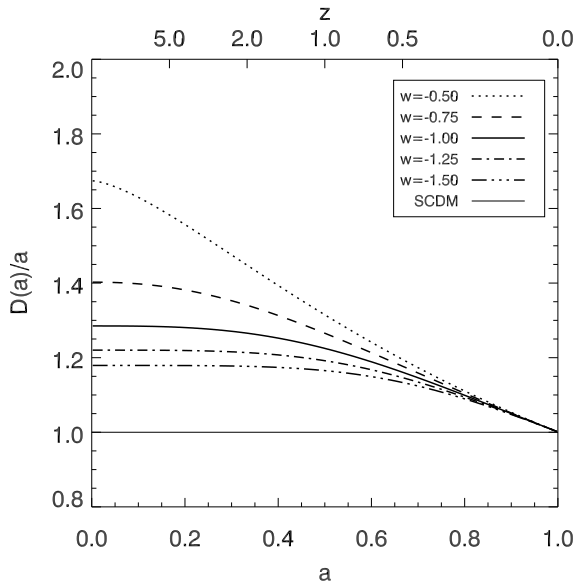}
\caption{
Ratio of the growth factor to the SCDM growth factor ($D_{\rm
SCDM}(a)=a$) as a function of scalefactor for different values of
$\wQ$. The growth factor is normalized to unity today ($D(a=1)=1$).
Linetypes are shown in the legend.
}
\label{fig:growthfactor}
\end{figure}

\subsection{The Spherical Collapse Approximation} 
\label{sec:tophat} 

A common convention is to define the virial mass and radius of a dark
matter halo by demanding that the mean density within the virial
radius of the halo be a factor $\Delta_{\rm vir}$ times larger than
the background density, $\rhob$.  Thus the virial mass and radius of a
halo are related by

\be 
\label{eq:Mvir} 
\Mvir=\f{4\pi}{3} \DeltaVir \rhob \Rvir^3.
\ee

In addition, the equivalent linear overdensity at collapse
$\delta_{\rm c}(z)$ is often used to determine the mass scale that is
typically collapsing at a given epoch.  Both of these quantities are
usually estimated using the approximation of spherical tophat collapse
(\eg Lacey \& Cole 1993).  In this section, we summarize the previous
results for spherical collapse in quintessence cosmologies (Mainini
\etal 2003; Mota \& van de Bruck 2004), and discuss the implications
for $\wQ < -1$ cosmologies.

The evolution of linear overdensities on scales much smaller than
those on which the quintessence field spatially clusters is

\be
\ddot \de +2H(a)\dot \de \simeq \f{3}{2}H_0\Om_M a^{-3}.   
\label{eq:growthfactor}
\ee
 
\noindent Solving this equation gives the growth factor for linear
perturbations.  Figure~\ref{fig:growthfactor} shows the linear growth
factor in five different quintessence models, normalized to the growth
factor in an $\Omega_{\rm M}=1$, SCDM cosmology in which $D(a) \propto
a$. Models with $\wQ \ge -1$ have been well-studied, with such models
showing more relative growth at higher redshifts. For models with $\wQ
\le -1$, figure~\ref{fig:growthfactor} shows how the trend towards
more relative growth at lower redshifts continues for $\wQ < -1$.

To detemine the non-linear growth of an object which has decoupled
from the expanding universe and virialized, we follow Wang \&
Steinhardt (1998) and Weinberg \& Kamionkowski (2003).  We use the
spherical collapse approximation to determine the non-linear
overdensity of a halo, $\DeltaVir(z)$, as a function of $\wQ$.  In
this model, an overdensity defined by $\DeltaVir(z)$ collapses at the
time $t_{\rm coll}$, when the radius of the overdense region
approaches zero. However, the actual, final radius of the collapsed
object is finite and can be computed using the virial theorem.  We
compute the equivalent linear overdensity at collapse by evolving the
linearized equation of motion, Eq.~(\ref{eq:growthfactor}), until the
time $t_{\rm coll}$, determined from the non-linear evolution of the
overdensity.  In the $\Lambda$CDM cosmology, the well-known results
for the linear and non-linear overdensities at collapse are
$\delta_c(z=0) \simeq 1.67$ and $\Delta_{\rm vir} \simeq 337$, and
vary with redshift.

\begin{figure}
\includegraphics[width=84mm]{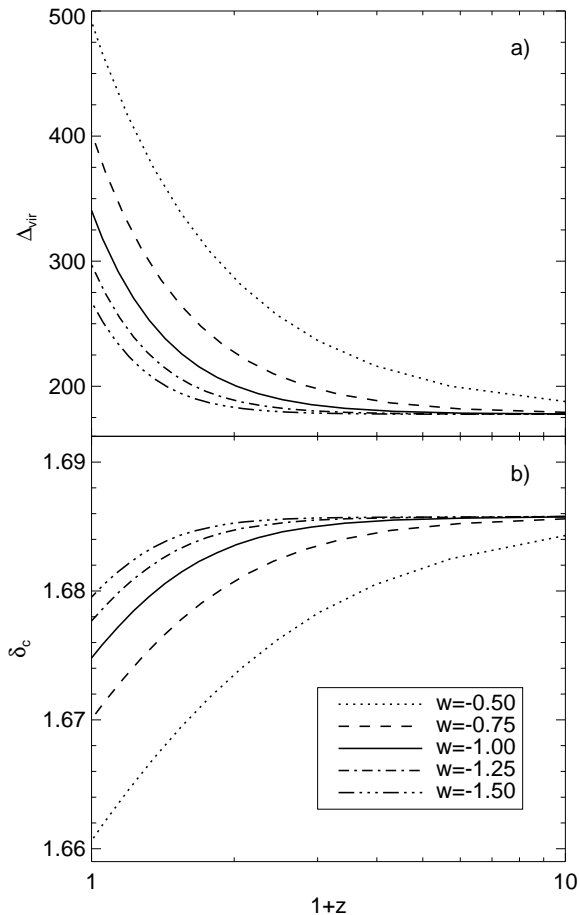}
\caption{
The non-linear and linear overdensities at collapse in quintessence
cosmologies.  In panel (a), we exhibit the non-linear overdensity at
collapse $\Delta_{\rm vir}(z)$ for 5 different quintessence models
that we explore in this paper ($\Omega_{\rm M}=0.3$).  The different
values of the quintessence equation of state parameter $\wQ$, are
shown in the legend.  The increase in $\Delta_{\rm vir}(z)$ for larger
$\wQ$ results primarily from later collapse redshifts.  In panel (b),
we show the equivalent linear overdensity at collapse $\delta_c(z)$,
as a function of redshift for the same quintessence models.
}
\label{fig:Dvir+dcoll}
\end{figure}

Figure \ref{fig:Dvir+dcoll} shows our results for the equivalent
linear overdensity at collapse $\delta_c(z)$, and the non-linear
overdensity at collapse $\Delta_{\rm vir}(z)$, as a function of
redshift in quintessence cosmologies.  The trend in $\delta_c(z)$
reflects the fact that over-densities grow more slowly in higher-$\wQ$
models.  We can understand the behavior of $\Delta_{\rm vir}{(w,z)}$
by noting that overdensities in models with larger $\wQ$ take longer
to collapse at late times.  Consider, for example, two haloes, one in
a $w=-1$ cosmology, and another in a $\wQ > -1$ cosmology, both of
which are just virializing at some redshift $z$.  Very roughly
speaking, the $w > -1$ halo will have had a turn-around time at a
higher redshift than the $w=-1$ halo, and its virial density will be
higher to reflect this.  Conversely $\wQ < -1$ models will have more
recent turn-around times and smaller virial overdensities.  We find
that an accurate fitting function, including the regime $\wQ < -1$,
can be obtained from a slight modification to the formula already
proposed by Weinberg \& Kamionkowski (2003) for $\wQ > -1$,

\be
\Delta_{\rm vir}(z) = 18 \pi^2 \left[ 1 + a \Theta^{b} (z) \right],
\label{eq:fittingfunction}
\ee

\noindent where $\Theta(z) \equiv \Omega_{\rm M}^{-1}(z)-1$, and with
$a = 0.432 - 2.001 ( \left |w \right |^{0.234} - 1)$ and $b = 0.929 -
0.222 (\left |w \right |^{0.727} - 1)$. We find this formula to be
accurate to better than $2\%$ for $0.1 \le \Omega_{\rm M} \le 1$ and
$-0.5 \le \wQ \le -1.5$.

\section{Analytic Model for Halo Concentrations}
\label{sec:halostruc}

\subsection{Main Ingredients}
\label{sec:B01model}

It is commonly agreed that the spherically-averaged density profiles
of dark matter haloes can be described fairly well by a generalized
NFW profile on scales that are resolved in state-of-the-art N-body
simulations:

\be
\rho_{\rm halo}(r) = \f{\rho_s}{(r/r_s)^\alpha(1-r/r_s)^{3-\alpha}},
\ee

\noindent where $\alpha$ describes the slope of the inner density
profile at $r<r_s$. The value of $\alpha$ that most closely represents
the results of N-body simulations is still debated, with acceptable
values between $-0.7$ and $-1.5$.  An additional complexity is that
recent studies indicate that haloes exhibit a range of inner slopes
(Klypin \etal 2001; Tasitsiomi \etal 2003; Navarro \etal 2003).  In
the following, we adopt $\alpha=-1$, corresponding to the standard NFW
profile. In this paper, we are concerned with the concentration
parameter $\cvir$, which is quite insensitive to the exact value of
$\alpha$ (see below).

The two parameters of the NFW profile are $r_s$ and $\rho_s$, with
$r_s$ the radius at which the logarithmic density slope becomes equal
to $-2$.  The concentration of the halo is defined as the ratio of its
virial radius to the scale radius of the NFW profile,

\be
\cvir=\f{\Rvir}{r_s}.
\ee

\noindent With these definitions, the halo virial mass is related to
the NFW parameters by

\be
\Mvir = 4 \pi \rho_s r_s^3 \left[ \ln(1+\cvir) - \f{\cvir}{1+\cvir} \right], 
\ee

\noindent so the halo density profile is completely determined by 
$\Mvir$ and $\cvir$.

Numerical simulations have revealed a correlation between $\Mvir$ and
$\cvir$, with halo concentrations log-normally distributed around the
median relation. Several simple models have been developed to explain
this correlation (NFW; B01; Eke, Navarro, \& Steinmetz 2001). Here we
focus on the B01 model and test the accuracy with which it predicts
the observed relation between halo mass and concentration in
simulations with $\wQ\ne-1$.

The B01 model assumes that a halo's central density and concentration
are set by the density of the universe at a characteristic formation
epoch.  This formation epoch qualitatively tracks the characteristic
collapse epoch for the halo subunits.  It is defined as the time when
the linear rms density fluctuations at a scale corresponding to a
fraction F of $\Mvir$ is equal to the linear collapse overdensity,
$\deltac$:

\be
\sigma(F \Mvir,a_c) = \deltac(a_c).
\label{eq:acoll}
\ee

Given the halo collapse epoch, the halo concentration is set via

\be
\cvir(\Mvir,a) = K_{\rm vir} \f{a}{a_c(\Mvir)}.
\label{eq:B01_vir}
\ee

A closely related study by Wechsler et al. (2002, hereafter W02)
showed that the B01 ``collapse epoch'' seems to correspond closely to
the epoch when the mass accretion rate of the halo
$\textrm{d}\ln(M_{\rm vir})/\textrm{d}t$, is large compared to the
rate of cosmic expansion.  W02 found that if one defines the end of
this rapid collapse phase to be when $\textrm{d}\ln(M_{\rm
vir})/\textrm{d}\ln(a) \le 2$, it corresponds closely to the
``formation epoch'' in Eq.~(\ref{eq:acoll}) above.  After this epoch
of rapid mass accretion ends, the halo mass and virial radius continue
to grow via comparably minor mergers and diffuse mass accretion. These
relatively minor mergers do not affect the inner regions of the halo
($r<r_s$) significantly and so $\Rvir$ grows, but $r_s$ remains
approximately constant, leading to an increase in concentration as the
halo evolves (W02).

The B01 model has two parameters, $F$ and $K$, that have to be
determined by calibrating them to numerical simulations. B01 analysed
two $\Lambda$CDM simulations with $\sigma_8=1.0$ and different
resolutions and box sizes.  Using a halo finder based on the Bound
Density Maxima (BDM) algorithm (Klypin \& Holtzman 1997), they
assembled a catalogue of several thousand haloes for each simulation,
and fit NFW profiles to each of them. B01 found that their model was
able to satisfactorily reproduce the mean relation between halo mass
and concentration, and the redshift dependence of the $\cvir$-$\Mvir$
relation with $F=0.01$ and $K=4.0$. B01 and W02 determined that the
scatter in concentration at fixed mass is well-described by a
log-normal distribution with $\sigc=0.14$~dex.  Other numerical
studies have found a somewhat smaller scatter, with Jing (2000)
reporting $\sigc=0.9-0.11$~dex, and Jing \& Suto (2002) finding
$\sigc=0.13$~dex.

The B01 model with $F=0.01$ and $K=4.0$ has since proven successful in
reproducing concentrations of $\Lambda$CDM haloes over more than six
orders of magnitude in halo mass from $M_{\rm vir} \simeq 10^7 \msun$
to $M_{\rm vir} \simeq 10^{13} \msun$ (\eg Col{\'{\i}}n et al. 2003;
Hayashi et al. 2003).  However, as discussed in B01, the model ceases
to make physical sense for halo masses large enough that $F \Mvir$
begins to approach the typical collapse mass at $z=0$.  This is
because linear fluctuations stop growing at late times in
$\Lambda$CDM, and with the simplified definition of collapse time
discussed above, very large haloes never collapse.  Consequently, the
B01 model with $F=0.01$ under-predicts halo concentrations for systems
more massive than a few $\times 10^{14} \msun$ (\eg Hayashi et
al. 2003; Dolag et al. 2003).  As a matter of pragmatism, this can be
remedied with a simple change of parameters.  With $F=0.001$ and
$K=3.0$ the model works adequately for all masses, though it becomes
somewhat less attractive because of the small value of $F$, for which
the collapse epoch no longer corresponds directly to $a_c$ defined by
W02.

\subsection {The Analytic Model in Quintessence Cosmologies}
\label{sec:B01Q}

As shown in \S~\ref{sec:Qstruc}, linear overdensities have greater
relative growth at higher redshift as $\wQ$ increases.  We then
expect, given an overdensity on a mass scale $\Mvir$, that this mass
scale will collapse at higher redshift as we increase $\wQ$.  As the
halo concentration reflects the density of the universe at the time of
rapid collapse, we expect this change in average formation time to
translate directly into a change in average concentration. The
differences in $\wQ \ne -1$ models should be confined to changes in
the rapid-collapse epoch $a_c$ for haloes of a given mass.

From Equation~(\ref{eq:acoll}), $a_c$ for a halo of mass $\Mvir$ is
determined by $\deltac$, $D(a)$, and $\sigma_8$. The changes in
$\deltac$ with $\wQ$ (Figure~\ref{fig:Dvir+dcoll}) have a very small
effect on $a_c$. More relevant are the changes in $\sigma(M,a)$ and
the linear growth rate (Fig.~\ref{fig:growthfactor}).  For a fixed
value of $\sigma_8$ all the $\wQ$-dependence of $a_c(\Mvir)$ will be
captured by $D(a)$, with $\sigma(M,a)$ reaching the $\deltac$ collapse
threshold at earlier times as $\wQ$ increases.  We therefore expect
concentrations to increase as $\wQ$ increases.

Note that an alternative definition of a halo's radius, $R_{\rm 200}$,
the radius at which the mean halo density is equal to $200 \rho_{\rm
crit}$ has frequently been used in the past.  This results in an
alternative definition of concentration: $c_{200} = R_{200}/r_s$
(e.g., NFW96).  Of course, given a set of cosmological parameters, the
model described above (and similar models) can be used to predict
equivalent relations between $c_{200}$ and $M_{200}$.  For a fixed
cosmology, the predicted relation between $c_{200}$ and $M_{200}$ will
look quite similar to the $\cvir$-$\Mvir$ relation, with an offset
that varies slowly as a function of concentration and accounts for the
differences in values of the outer halo radius.

Because the simulations of B01 focused only on one cosmology, it was
impossible to tell whether agreement with the simulations and the
proposed B01 model was sensitive to the choice of defining halo
concentration relative to $R_{\rm vir}$ instead of $R_{200}$. That is,
one could have equally well proposed a different model based on
$c_{200}$:

\be
c_{200}(M_{200},a) = \tilde{K}_{200} \f{a}{a_c(M_{200})}.
\label{eq:B01_200}
\ee

\noindent The simulation results of B01 could have been reproduced
using this model, simply by setting $\tilde{K}_{200}$ equal to a
slightly smaller value than the original $K=4.0$.

Consider now the current case, where we compare simulation results
based on cosmologies with different $w$'s and hence different
$\Delta_{\rm vir}$ values.  In these simulations, the ratios of
$R_{\rm vir}/R_{200}$ (and $\cvir/c_{200}$) for fixed-mass haloes will
depend on the value $\Delta_{\rm vir}(w)$.  Therefore, it is
impossible for a model based on $c_{200}$ with fixed $\tilde{K}_{200}$
to do equally well as the original B01 model based on $\cvir$ with
fixed $K$ for all values of $w$.  More physically, the original B01
model described above implicitly \textit{assumes} that the haloes have
virialized at the appropriate virial density and predicts that, in
addition, the halo collapse redshift acts to set the {\textit ratio}
of the virial density ($\propto \Rvir^{-3}$) to the central density
($\propto r_s^{-3}$).  A model based on $c_{200}$ with fixed
$\tilde{K}_{200}$ would instead assume that all of the changes in halo
density arise solely because of changes in $a_c$.  As we demonstrate
below, the virial assumption seems to capture better our simulation
results.  It seems therefore, that there are two physical processes
that set halo densities: one process is related to the global process
of halo virialization and the other may be related to an earlier,
rapid-collapse epoch.

\section{Numerical Simulations}
\label{sec:sims}

In this section, we describe our numerical simulations.  In
\S~\ref{sec:params}, we detail the numerical and cosmological
parameters that were used. In \S~\ref{sec:halofinder}, we describe the
methods we use to locate haloes and to fit NFW profiles to their
density profiles.

\subsection{Simulations and Parameters}
\label{sec:params}

We use GADGET version 1.1, a publicly-available and well-tested
$N$-body code (Springel \etal 2001).  Gravity between particles is
solved using a hierarchical tree algorithm in co-moving coordinates,
and both the force calculations and the time-stepping are performed in
a fully adaptive way. Using the parallel version, we have run the code
on either 96 375MHz IBM Power3 processors of NERSC's \textit{Seaborg}
or on 64 1.4GHz Athlon processors of \textit{UpsAnd}, a 264-processor
Beowulf cluster at The University of California at Santa Cruz.  We
made necessary alterations to the expansion rate of the universe for
GADGET to account for quintessence cosmologies with $\wQ \ne -1$.

Power \etal (2003) have performed a detailed convergence study of a
high resolution cluster simulation using GADGET, and although we
simulate a much larger cosmological volume we have followed their
recommendations for a number of GADGET's parameters. In particular we
have chosen an adaptive timestep equal to $\Delta t_i=\eta_{a\epsilon}
\sqrt{\epsilon_i/a_i}$, where $\epsilon_i$ and $a_i$ are the
gravitational softening and acceleration experienced by the $i^{\rm
th}$ particle in the simulation, and $\eta_{a\epsilon}$ is a
dimensionless constant. Power \etal (2003) recommend setting
$\eta_{a\epsilon}=0.2$; this choice of adaptive timestep minimizes
undesirable effects due to particle discreteness and hard scatterings,
while at the same time allowing for convergence at minimal
computational expense.  In GADGET, gravitational softening is
performed using a cubic spline (Springel \etal 2001), for which the
potential becomes exactly Newtonian at $r=2.8~\epsilon$, where
$\epsilon$ is the softening length. Generally our simulations were run
with a co-moving softening length of $\epsilon=2.5 \hinv$~kpc,
although we have run a few cases with $\epsilon$ as low as
$1\hinv$~kpc.

Our cosmological background model is fixed by $\Om_M=0.3$,
$\Omega_{\rm Q}=0.7$, $h=0.7$, and $n=1.0$ for all values of $\wQ$.
In normalizing $\sigma_8$ on the scale of galaxy clusters, the initial
power spectra are nearly unaffected by quintessence. However, when
normalizing to the scales probed by the CMB, the initial power spectra
are altered by the inclusion of quintessence
(section~\ref{sec:norm}). As discussed below, we normalize our
simulations such that $\sigma_8 \simeq 0.74$, thus the effect of $\wQ
\ne -1$ is due almost exclusively to the expansion rate.

All of our simulations were run with $256^3$ particles in boxes with
sides of length $60\hinv$Mpc. $\Om_M=0.3$ implies a mass per particle
of $M_p=1.1 \times 10^9 \hinv \msun$. For the analysis of halo
concentrations we used only haloes with more than 100 particles (see
\S\ref{sec:concentration}). This corresponds to a minimum halo mass of
$\Mvir^{\rm min}\simeq 1.1 \times 10^{11} \hinv \msun$, for which the
B01 model predicts a median concentration of 13.5 (for
$\sigma_8=0.74$). This translates into an NFW scale radius of
$r_s^{\rm min} \sim 10 \hinv$~kpc. More massive haloes will have
larger scale radii, and because even this minimum scale radius is
almost three times larger than our softening length, we should be able
to determine accurate concentrations from the haloes in our
simulations.

Table \ref{tab:parameters} summarizes the parameters used in our
simulations.

\begin{table}
\caption{Simulation Parameters and Power Spectrum Normalizations.}
\label{tab:parameters}
\begin{center}
\begin{tabular}{cccc}
\hline
Model $\wQ$ & $\epsilon/$h$^{-1}$ kpc & ${\sigma_{8,{\rm nom}}}^a$ & ${\sigma_{8,{\rm eff}}}^a$ \\
\hline
$-0.50$ & $2.5$ & $0.742$ & $0.799$ \\
$-0.75$ & $2.5$ & $0.740$ & $0.775$ \\
$-1.00$ & $2.5$ & $0.738$ & $0.716$ \\
$-1.00$ & $1.0$ & $1.000$ & $0.972$ \\
$-1.25$ & $2.5$ & $0.736$ & $0.716$ \\
$-1.50$ & $2.5$ & $0.734$ & $0.714$ \\
\hline
\end{tabular}
\end{center}

\medskip

All other parameters are fixed at the same value for all simulations.
The number of particles is $N_p=256^3$, the box size is $L_{\rm
box}=60h^{-1}{\rm Mpc}$, and the initial redshift is $z_i=50$. For all
simulations, the remaining cosmological parameters are $\Omega_{\rm
M}=0.3$, $\Omega_{\rm Q}=0.7$, $h=0.7$, and $n=1.0$.\\ $^a$ the
difference between $\sigma_{8,{\rm nom}}$ and $\sigma_{8,{\rm eff}}$
is explained in \S\ref{sec:initialcond}.
\end{table}

\subsection{Initial Conditions}
\label{sec:initialcond}

Setting initial conditions for our simulations requires fixing the
$z=0$ power spectrum normalization, which we parameterize by
$\sigma_8$.  As discussed above (\S\ref{sec:norm}), our current
understanding of present day clusters of galaxies makes $\sigma_8$
still uncertain by $\sim 20-30$\% ($\sigma_8 \sim 0.70-1.10$ for
$\Omega_{\rm M}=0.3$). Halo concentrations in cosmological N-body
simulations depend sensitively on $\sigma_8$, because the amount of
small-scale power directly affects typical halo formation times,
especially for high mass haloes. For example, at $\Mvir=10^{14}\ \hinv
\msun$ the B01 model predicts median concentrations of $\cvir = 6.8$
and $\cvir = 5.3$ for $\sigma_8=0.90$ and $\sigma_8=0.74$,
respectively.  We initialize our simulations with the values of
$\sigma_8$ determined by Kuhlen \etal (2004) from the abundance of
clusters in the HIFLUGCS sample of local clusters (Reiprich \&
B\"ohringer 2002): $\sigma_8=0.742, 0.740, 0.738, 0.736, 0.734$ for
$\wQ=-1.50, -1.25, -1.00, -0.75, -0.50$, respectively.  We construct
initial conditions for each $\wQ$ with the routines of the
publicly-available PM code (Klypin \& Holtzman 1997).

\begin{figure}
\includegraphics[width=84mm]{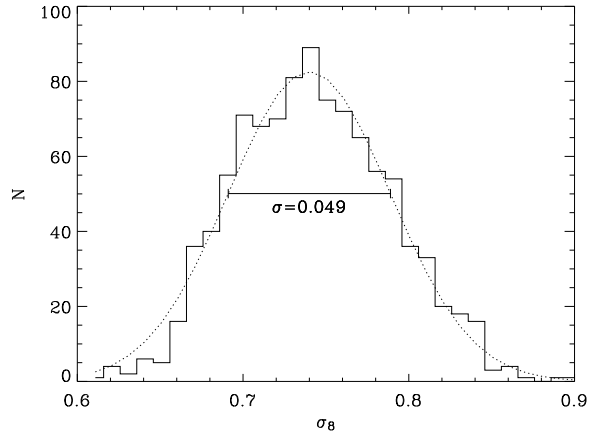}
\caption{
A histogram of the effective values of $\sigma_8$ determined from 1000
Monte-Carlo realizations of the initial conditions with an input
powerspectrum normalized to $\sigma_8=0.74$. The distribution is
Gaussian with a standard deviation of $0.049$, or roughly $7\%$.
}
\label{fig:sigma8_mc1000}
\end{figure}

The process of initializing particle positions and velocities based on
the linear power spectrum is subject to statistical fluctuations. The
largest modes of the system ($\lambda \simeq \Lbox$) are sampled only
twice per dimension and are thus sensitive to deviations caused by
small number statistics. As $8 \hinv$~Mpc is close to $\Lbox$, this
can lead to noticeable differences between the ``nominal'' value of
$\sigma_{8,{\rm nom}}$, used to construct the initial conditions and
an ``effective'' value of $\sigma_{8,{\rm eff}}$, determined from the
actual particle positions at the initial redshift. In order to
quantify this difference, we determine $\sigma_{8,{\rm eff}}$ by
direct numerical integration of the initial $N$-body power spectrum:

\be
\sigma_{8,{\rm eff}}^2 = \f{1}{2\pi^2} \f{D^{2}(z=0)}{D^{2}(z=50)}
\int_0^\infty k^2 P_{\rm
  num}(k) W(kR_8)^2 \textrm{d}k,
\label{eq:sigma8direct}
\ee

\noindent where $P_{\rm num}(k)$ is the power spectrum derived from
the N-body initial conditions and $W(x)$ is the spherical tophat
window function given by $W(x)=3/x^2 (\sin x / x - \cos x)$, evaluated
at $kR_8$, with $R_8=8\hinv$~Mpc.

$P_{\rm num}(k)$ only extends to $k_{\min}\approx 0.1 \hinv {\rm
Mpc}^{-1}$, which is not low enough to allow the integral in
Equation~\ref{eq:sigma8direct} to converge. We estimated the portion
of the integral below $k_{\rm min}$ by integrating the smooth
analytical power spectrum, and applied this correction to get
$\sigma_{8,{\rm eff}}$. We found that for $\wQ \leq -1$
$\sigma_{8,{\rm eff}} \simeq 0.715$, which is $\sim 3 \%$ lower than
$\sigma_{8,{\rm nom}}$.  However, the variation of $\sigma_8$ in a box
of this size due to cosmic variance should be roughly $\sim 5\%$, so
this difference is not surprising.  To demonstrate this explicitly, we
have constructed $1000$ realizations of the initial conditions and
computed values of $\sigma_8$.  We infer from these realizations that
measured values of the effective $\sigma_8$ are distributed with a
standard deviation of $\sim 5\%$. The resulting distribution is shown
in Fig.~\ref{fig:sigma8_mc1000}.  Therefore, the difference between
$\sigma_{8,{\rm eff}}$ and $\sigma_{8,{\rm nom}}$ for the three $w \le
-1$ cases is not surprising and is consistent with cosmic variance.
We find similar deviations of $\sigma_{8,{\rm eff}}$ from
$\sigma_{8,{\rm nom}}$ for all of the values of $\wQ$ that we
simulate.

\subsection{Halo Finders}
\label{sec:halofinder}

We use two different halo finding algorithms to locate the haloes in
our simulations, depending upon the quantities we probe with our
simulation data.  In \S~\ref{sec:mf}, we compare the mass functions of
our simulated haloes with the J01 ``universal'' mass function.  J01
used the Friends-Of-Friends (FoF) algorithm (Davis \etal 1985) to
identify simulated haloes.  In order to make a direct comparison to
the J01 mass functions, we have employed a University of Washington
FoF halo finder (\texttt{\small
http://www-hpcc.astro.washington.edu/tools/fof.html}).  As in J01 we
set the linking length to 0.2 times the mean inter-particle separation
for all models.

To directly compare our halo concentrations to the B01 model
(\S~\ref{sec:concentration}), we use an updated version of the halo
finder that B01 and W02 employed in order to identify haloes.  This
halo finder is based on the BDM algorithm (Klypin \& Holtzman 1997)
and iteratively removes particles that are not bound to the halo in
question.  Upon identifying haloes, we fit NFW profiles to each halo
and determine $\cvir$. For more detail, we refer the reader to
Appendix B of B01 and Appendix A of W02.  We include in our catalogues
only haloes with more than $100$ particles, the same cut-off used by
W02.

We have checked that both halo finding algorithms agree with each
other, within our expectations, by comparing mass functions. The
systematic differences in total mass between haloes defined in terms
of the cosmology-dependent virial overdensity $\DeltaVir(w)$ (as in
the BDM finder) and those based on the fixed FoF linking length
translate into only minor differences in mass functions.  At high
redshift, the two mass functions actually converge because
$\DeltaVir(z)$ approaches $178$ and a linking length of $0.2$ times
the mean inter-particle separation roughly corresponds to a mean halo
density of $\sim 180 - 200$ times the background density.  Below $z
\sim 2-2.5$, the FoF and BDM mass functions agree well.  At higher
redshifts the BDM-based finder becomes increasingly incomplete. A
consequence of the low value of $\sigma_8$ in our simulations is that
halo formation occurs more recently. Frequent merger events during the
rapid mass growth phase of halo formation disrupt any spherical
symmetry in the halo density profile.  These haloes will not be well
described by the NFW formula. In our implementation of the BDM halo
finder, haloes with very bad fits to the NFW profile are rejected and
not included in the catalogues. This is the major source of
incompleteness at $z \ga 2.5$.  Note that the net effect of this
incompleteness is to underestimate the number of low concentration
haloes at high redshift.

\section{Results}
\label{sec:nbody_results}

\subsection{Mass Functions}
\label{sec:mf}

\begin{figure*}
\includegraphics[width=170mm]{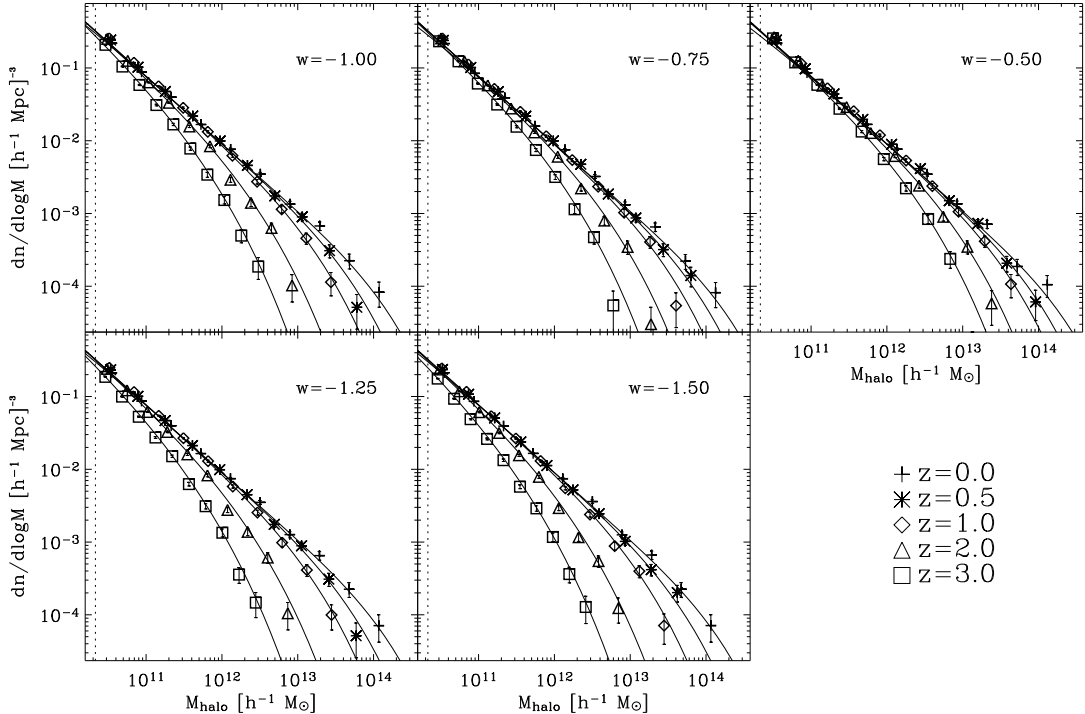}
\caption{
A comparison between the J01 analytic mass function (solid lines) and
the mass functions derived from our five N-body simulations (shapes
with error bars, see the legend in the lower right portion of the
Figure).  The haloes in our simulations were located using a FoF
algorithm with a linking length equal to 0.2 times the mean particle
separation as in J01.  In each panel we plot mass functions at various
redshifts, from top to bottom: $z = 0.0, 0.5, 1.0, 2.0, \textrm{ and }
3.0$. The error bars represent the Poisson noise due to the finite
number of haloes in each mass bin.
}
\label{fig:mf_allw}
\end{figure*}

Several recent numerical studies have demonstrated that the J01
formula for halo mass functions may be considered ``universal'' as it
accurately describes halo counts as a function of mass in $N$-body
simulations of various cosmologies, including models containing dark
energy with $\wQ \ne -1$ and a time-varying $\wQ$ (Linder \& Jenkins
2003; Klypin \etal 2003; Macci\`o \etal 2003; \L okas, Bode, \&
Hoffman 2003).  We confirm and further extend this conclusion by
presenting halo mass functions from our simulations at different
redshifts.  In Figure~\ref{fig:mf_allw}, we show the mass functions of
our FoF haloes in each cosmology and at a variety of redshifts from
$z=0$ to $z=3$.  It is apparent that, in all panels, the mass
functions and the redshift evolution of the mass functions for each
$\wQ$ model are in excellent agreement with the J01 forumula.  Thus we
confirm that the J01 formula is a good approximation at all redshifts
for $\wQ > -1$ and we extend the range of validity of the J01 relation
to include quintessence models with $\wQ < -1$.

\subsection{Concentrations}
\label{sec:concentration}

In \S\ref{sec:B01Q}, we described the manner in which quintessence
modifies the predictions of the analytic B01 model for halo
concentration as a function of mass.  We have assembled catalogues
consisting of more than $\sim 1600$ haloes for each of our
quintessence $N$-body simulations. The density profiles of every halo
have been fit to an NFW profile, yielding a best-fit $\cvir$ for each
object.  Some of these fits produced concentrations smaller than one.
We have excluded these haloes from our subsequent analysis.

\begin{figure*}
\includegraphics[width=170mm]{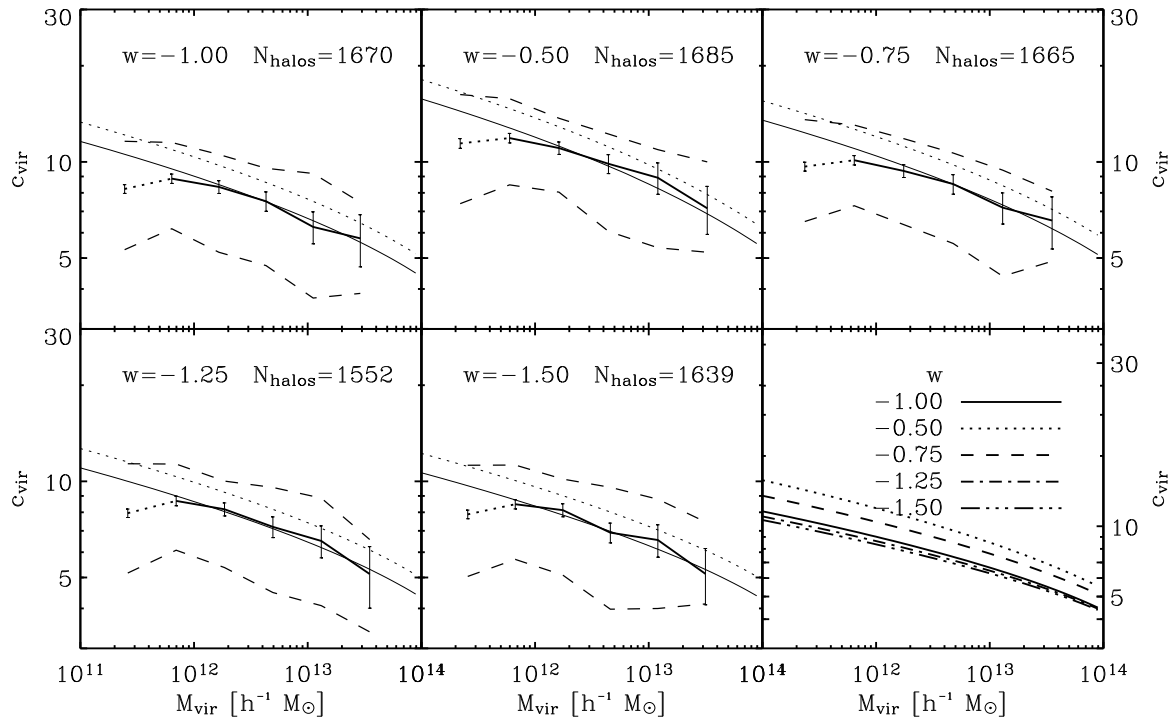}
\caption{
The concentration parameter $\cvir$, as a function of mass $\Mvir$,
for the five quintessence N-body simulations. The original ($F=0.01,
K=4.0$) and our best-fitting ($F=0.01, K=3.5$) B01 models are
over-plotted as thin dotted and solid lines, respectively. Error bars
represent the Poisson noise due to the finite number of haloes per
bin. The dashed lines are our estimates of the intrinsic scatter in
the relation, obtained by removing the scatter due to errors in the
fits of the NFW profiles, as well as Poisson noise. The lowest mass
bin was excluded in our analysis and is shown here for completeness
only. The total number of haloes in the remaining bins is shown in the
upper right corner of each panel.  The lower right panel shows the
best-fitting B01 model for all five values of $\wQ$.
}
\label{fig:cofM}
\end{figure*}

The resulting $c_{\rm vir}(M_{\rm vir})$ relations are plotted in
Figure~\ref{fig:cofM}.  We find that our simulations produce haloes
with slightly lower concentrations than expected from previous
simulation results (\eg B01, Col{\'{\i}}n et al. 2003) and the
analytic model proposed by B01 with $F=0.01$ and $K=4.0$.  However, we
find that this difference can be described well by a constant offset.
For example, keeping $F=0.01$ and lowering the proportionality
constant $K$ to $K = 3.5$ [see Eq.~\ref{eq:B01_vir}] in the B01 model
matches our data quite well for all of the $\wQ$ models we explore.
We discuss this overall offset further in \S~\ref{sec:GADGETvsART}.
The $c_{\rm vir}$-$M_{\rm vir}$ relation flattens out below $\Mvir
\approx 6 \times 10^{11}\hinv\msun$.  We attribute this to the lower
number of particles in these haloes, making them more susceptible to
relaxation effects which tend to cause the central regions of haloes
to be more diffuse and lead to lower concentrations. It is thus
unlikely that this flattening represents any physical effect (compare
to the results of Col{\'{\i}}n et al. 2003), and we have neglected the
lowest mass bin in determining the best-fit value of $K$.

As mentioned above (\S\ref{sec:B01Q}), considering several
cosmological models with different virial overdensities $\DeltaVir$
allows us to distinguish between analytic prescriptions based on
definitions of halo concentration in terms of $R_{200}$, in which the
proportionality constant $\tilde{K}_{200}$ is independent of cosmology
(Eq.~\ref{eq:B01_200}), and those based on the virial radius $R_{\rm
vir}$, in which $K_{\rm vir}$ is cosmology-independent
(Eq.~\ref{eq:B01_vir}). To test this, we re-analysed the five $z=0$
$N$-body outputs using the BDM halo finder, but setting
$\DeltaVir=\rho_{\rm vir}/\rhob=\rho_{\rm vir}/\rho_{\rm crit} \;
\Omega_M^{-1}=200 \; \Omega_M^{-1} \simeq 667$, effectively yielding a
relation between $c_{200}$ and $M_{200}$. Matching these relations to
the model described by Eq.~\ref{eq:B01_200} we determined best-fitting
values of $\tilde{K}_{\rm 200}= (3.76, 3.44, 3.32, 3.16, 3.16)$ for
$\wQ=(-0.50, -0.75, -1.00, -1.25, -1.50)$, respecively. This range in
$\tilde{K}_{200}$ is not consistent with one cosmology-independent
value of $\tilde{K}_{200}$. The results of this analysis suggest that
models similar to the B01 model, in which the halo concentration is
defined in terms of $R_{\rm vir}$ and $\DeltaVir$, are more readily
generalizable to alternative cosmologies as $\cvir$ is related to
$a/a_c$ via a \textit{cosmology-independent} constant of
proportionality, $K_{\rm vir}$.  Put another way, defining the radius
of a halo, and thus its concentration, using a fixed overdensity
criterion necessitates using a cosmology-dependent proportionality
constant in Eq. (\ref{eq:B01_vir}) while the cosmology-dependent
virial overdensity definition seems to account for these differences,
so that $K_{\rm vir}$ is independent of cosmology.

As in previous studies (B01; Jing 2000; Jing \& Suto 2002), we also
find that haloes of a given mass have a broad distribution of
concentrations. To determine the inherent scatter in the
$\cvir$-$\Mvir$ relation it is important to account for the artificial
scatter introduced by uncertainties in the fit to an NFW profile and
by the Poisson noise in each bin.  Following the B01 analysis, we
corrected for the former by determining 500 one-sided Gaussian
deviates for each halo with a standard deviation equal to the error in
the $\cvir$ fit returned by the halo finder. The deviates are positive
(negative) if $\cvir$ is less (greater) than the median in that
bin. We then determined the $16^{th}$ and $84^{th}$ percentiles in
$\log(c_{\rm vir})$ and subtract off the Poisson noise from each in
quadrature. The resulting estimates of the intrinsic scatter are shown
as the dashed lines in Figure~\ref{fig:cofM}. The scatter is
consistent with being independent of $\wQ$ and $\Mvir$, and we find
that taking the B01 proportionality constant to be $K_{\rm low}=2.28$
and $K_{\rm high}=4.52$ fits the lower and upper lines well. These
values correspond to $\sigc$$_{,\rm low}=0.18$~dex and $\sigc$$_{,\rm
high}=0.11$~dex. Although these are similar in magnitude to the
scatter reported in previous studies, our distributions are skewed
away from log-normal toward lower concentrations.  We note that the
skewness may likely be caused by the lower resolution of our
simulations, which tends to result in lower concentration haloes.

\begin{figure*}
\includegraphics[width=170mm]{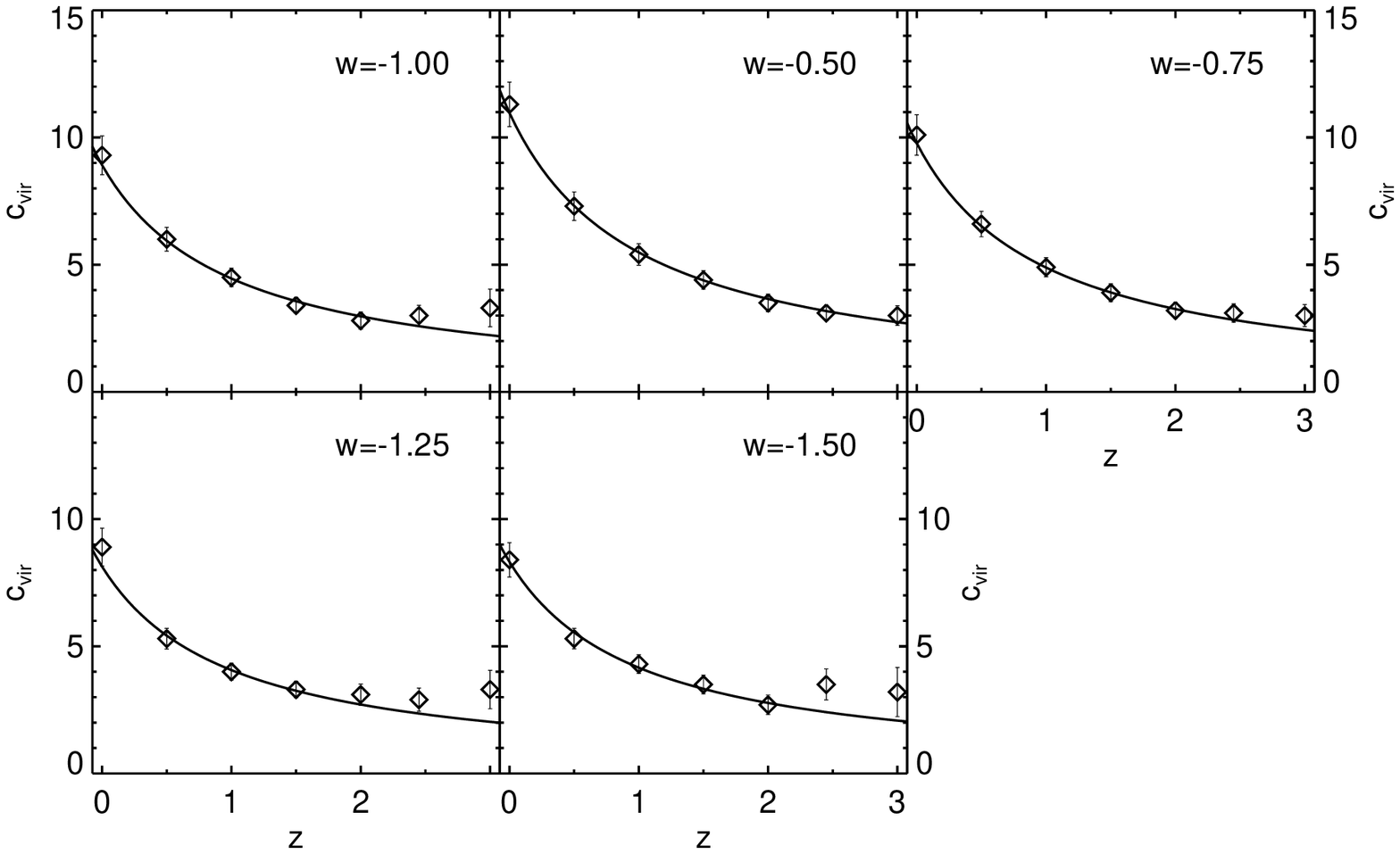}
\caption{
The median concentrations of haloes in bins of mass $\Mvir = 7 \times
10^{11} \hinv \msun \pm 0.05$ dex as a function of redshift. Error
bars represent the Poisson noise due to the finite number of haloes
per bin. $\cvir$ falls in proportion to $1/(1+z)$ (solid lines), in
agreement with the B01 model. The flattening at $z\ga2.5$ is probably
not physical, and may be caused by incompleteness in the halo
catalogues at high redshift which preferentially excludes low
concentration haloes.
}
\label{fig:cofz}
\end{figure*}

For a fixed mass the B01 model predicts that concentration should
decrease with redshift as $1/(1+z)$. The haloes in our simulation also
satisfy this relation, as shown by Figure~\ref{fig:cofz}, in which we
plot the redshift dependence of concentration for haloes of mass
$\Mvir=7 \times 10^{11} \hinv\msun$. This figure shows that the
concentrations follow the $c_{\rm vir} \propto (1+z)^{-1}$ relation
that is embodied in the B01 analytic model.  At redshifts greater than
$\sim 2.5$, our catalogues of haloes in this mass bin with fitted NFW
profiles become incomplete. This incompleteness preferentially affects
low concentrations haloes, causing the $\cvir(z)$ relation to flatten
at high redshift.  We do not believe this to be a physical effect, and
trust our data points to $z \sim 2.5$.

\subsection{The Concentration Discrepancy}
\label{sec:GADGETvsART}

We have attempted to understand the origin of the discrepancy between
$\cvir(\Mvir)$ derived from our simulations and those reported by B01
and summarized by the B01 model.  We have re-analysed the same $z=0$
simulation data that was analysed previously by B01, and we were able
to reproduce their $c_{\rm vir}$-$M_{\rm vir}$ relation and
scatter. We conclude that the discrepancy that we observe is not due
to any change in analysis procedures.

\begin{figure}
\includegraphics[width=84mm]{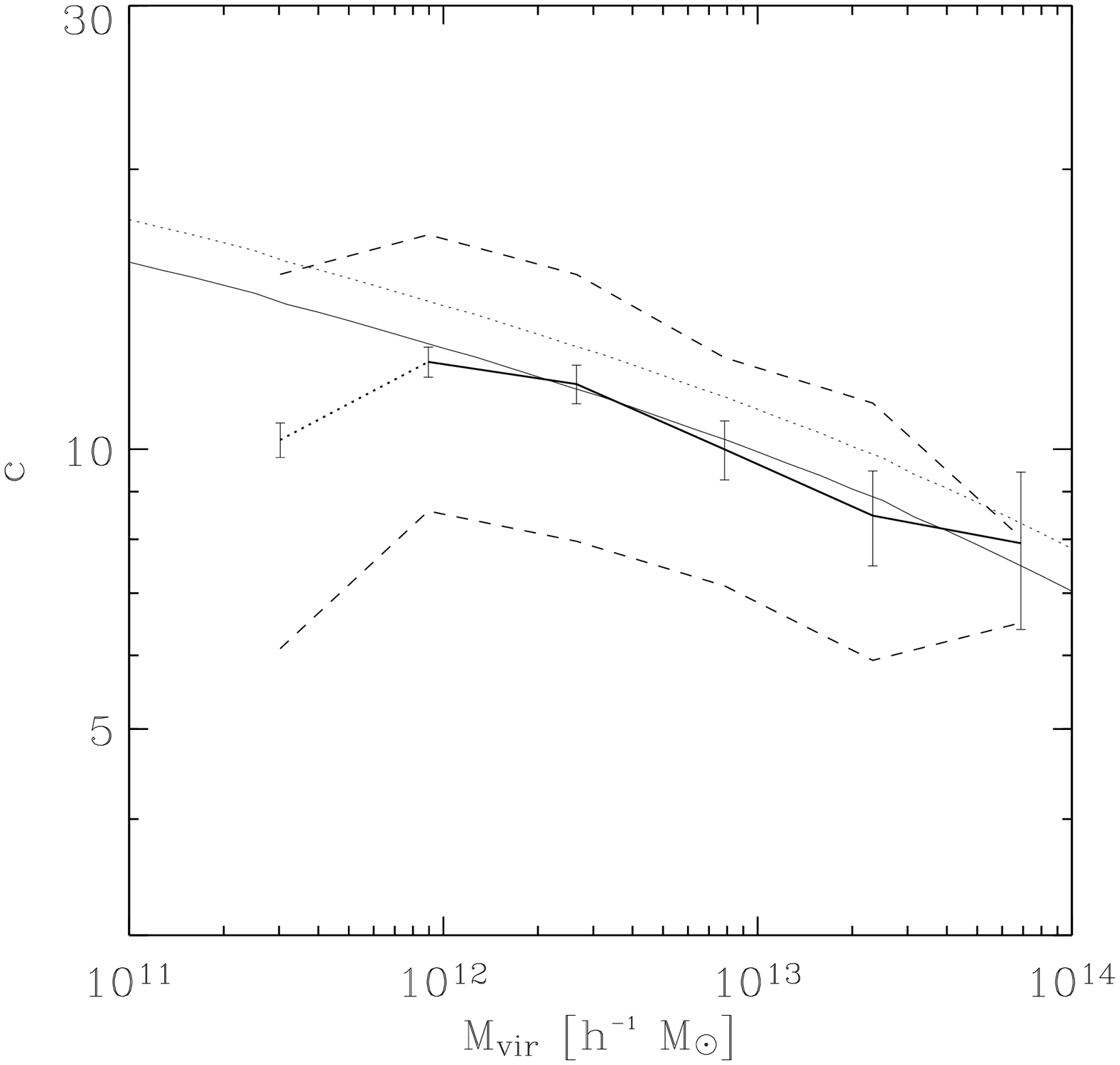}
\caption{
$\cvir(\Mvir)$ for our high resolution ($\epsilon=1.0 \hinv$~kpc)
$\sigma_8=1.0$ GADGET simulation designed to approximate the
adaptive-refinement ART simulation used in B01. The original B01
relation is overplotted as a dotted line.  The dotted line segment and
the dashed lines are as in Fig. \ref{fig:cofM}.
}
\label{fig:GADGEThires}
\end{figure}

Of course, the main difference between the study of B01 and our work
is the choice of simulation codes. Whereas we use the
publicly-available, uniform-resolution code GADGET, B01 used the
adaptive-refinement code ART.  Undoubtedly the effective resolution at
the centres of haloes was higher in the B01 simulation than in ours.
In order to shed further light on this matter, we have run one
additional GADGET simulation designed to test the importance of the
effective force resolution. Compared to the five simulations discussed
previously, this one has higher force resolution ($\epsilon=1.0
\hinv$~kpc) and $\sigma_8=1.0$.  The resulting $\cvir(\Mvir)$ relation
is shown in Figure~\ref{fig:GADGEThires}.

Here, we again find that the GADGET concentrations are systematically
lower than the ones found by B01 with ART by $\sim 14\%$. Instead of
$K=4.0$ we find $K=3.44$ matches the GADGET halo concentrations. This
is consistent with the value found for the five lower resolution
quintessence simulations described above. The difference in $K$
between our simulations and the B01 simulation may be due to an
inherent difference between the GADGET and ART $N$-body codes. Whether
this is due to the higher maximum force resolution afforded by the
adaptive refinement of ART, or another difference between the two
codes remains unclear.

Several recent analyses based on $w=-1$ simulations with higher
resolution than our own (Hayashi et al. 2003; Col{\'{\i}}n et
al. 2003; Tasitsiomi et al.  2003) also favor the B01 model with
$K=4.0$.  In light of these results we suggest that our GADGET
simulations systematically under-predict halo concentrations by $\sim
10-15\%$.  However, when this offset is normalized out, the variation
of $\cvir(\Mvir)$ with $\wQ$ scales as predicted, and we conclude that
the B01 model is successful in this regard.  Emboldened by this
success, we use the model to explore the implications of various
normalization choices and to compare expected halo densities with
those inferred from galaxy rotation curves.  In what follows, we
assume $K=4.0$ for the model normalization and we advocate this choice
for the reasons outlined above.  (However, setting $K=3.5$ would not
qualitatively change the conclusions that follow.)

\section{Halo concentrations and central densities with $\wQ \ne -1$}
\label{sec:DeltaVby2}

\begin{figure*}
\includegraphics[width=128mm]{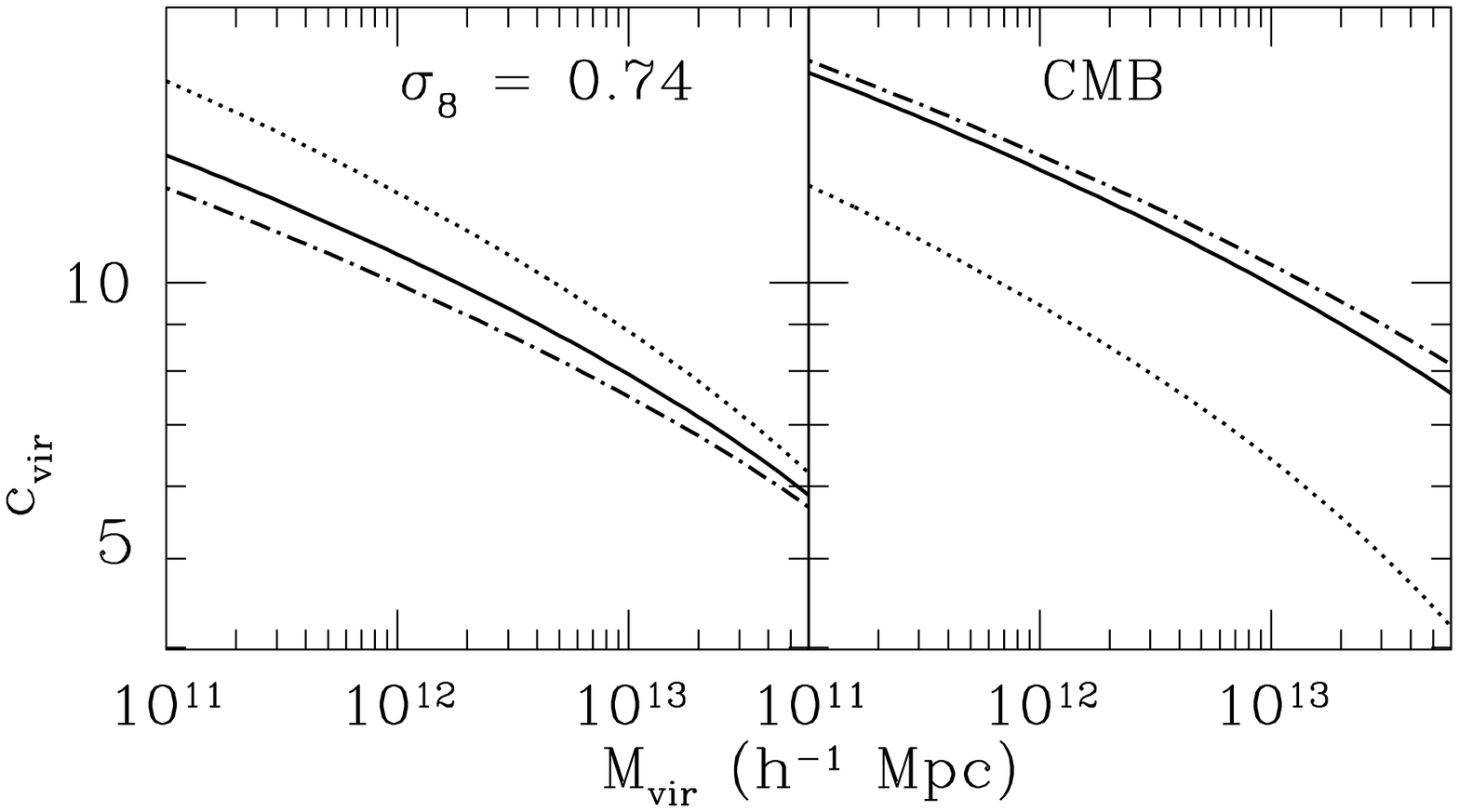}
\caption{   
Predictions for halo concentration as a function of $M_{\rm vir}$.
Dotted, solid, and dash-dot lines in each panel refer to $\wQ =
-0.5$,$-1$, and $-1.5$, respectively.  The left panel is for the fixed
$\sigma_8$ used in the simulations presented in this paper. The right
panel shows how the halo concentration changes when we normalize based
on the CMB: $\sigma_8 \simeq 0.6$ , $0.9$, and $1.0$ for $\wQ =
-0.5$,$-1$, and $-1.5$, respectively.
}  
\label{fig:cvirplot}
\end{figure*}

In Figure~\ref{fig:cvirplot}, we illustrate the degeneracy between
$\wQ$ and $\sigma_8$ in setting halo concentrations.  Shown are the
predictions of the B01 model ($F=0.01$, $K=4.0$) for $\cvir (z=0)$
with a fixed normalization $\sigma_8=0.74$ for several values of $w$.
As discussed in \S~3.2, for a fixed normalization, concentrations
increase as $\wQ$ increases because haloes collapse earlier.  The
right panel of the figure shows the corresponding predictions for
$\cvir$ with $\sigma_8$ determined by normalizing to the CMB with
$\tau = 0$ (see \S~2.3).  Note that for the CMB normalization, the
trend with $\wQ$ is in the opposite direction, with increasing $\wQ$
implying lower concentrations.  As illustrated in Figure
\ref{fig:sig8vswQ}, higher $\wQ$ requires a lower normalization:
$\sigma_8 \simeq 0.6$, $0.9$, and $1.0$ for $\wQ = -0.5$, $-1$, and
$-1.5$, respectively.  The change in normalization based on the CMB
dominates changes in the growth function that give rise to the
behavior of $\cvir(w)$ at fixed normalization.

As discussed previously, the $\cvir$ parameter is useful, but it is
not a direct measure of physical density.  We would like, therefore to
convert our predicted $\cvir$ relations into quantities that have a
more direct physical interpretation, and are more amenable to
comparison with observations.  Alam, Bullock, \& Weinberg (2002)
proposed the {\em central density parameter} as a means to quantify
the physical density in the central regions of a galaxy:

\be
\label{eq:DVby2} 
\Delta_{\rm V/2} \equiv 
\f{1}{2} \left( \f{{\rm V}_{\rm max}}{H_{0} r_{{\rm V}/2}} \right)^{2}. 
\ee

\noindent $\Delta_{\rm V/2}$ is the mean overdensity within $r_{\rm
V/2}$, the radius at which the galaxy rotation curve reaches half its
maximum, $V_{\rm max}$.

The $\Delta_{\rm V/2}$ parameter is advantageous for several reasons.
First, it facilitates comparisons between theory and observation.  Any
predicted $\cvir$ vs. $\Mvir$ relation can be easily converted into a
$\Delta_{\rm V/2}$ vs.  $V_{\rm max}$ relation.  Similarly, given an
observed galaxy rotation curve, $\Delta_{\rm V/2}$ can be determined
without reference to any particular analytic density profile.  It also
has the useful characteristic that even if an observed rotation curve
is rising at the last measured point, substituting the highest (outer)
most point on the rotation curve for $V_{\rm max}$ in the formula
above results in an {\em upper limit} on the true value of
$\Delta_{\rm V/2}$.  Specifically, if $V_{\rm max}$ is underestimated
by a factor $f_{\rm v}$, and the density profile varies with radius as
$\rho(r) \propto r^{-\alpha}$, this leads to an overestimate of
$\Delta_{\rm V/2}$ by a factor of $f_{\rm v}^{-2\alpha/(2-\alpha)}$.
Thus this is an overestimate so long as the density profile falls off
with radius or is constant.  Furthermore, if $\alpha > 2/3$, the
fractional overestimate of $\Delta_{\rm V/2}$ is larger than the
fractional underestimate of $V_{\rm max}$.

\begin{figure*}
\centering 
\includegraphics[width=128mm]{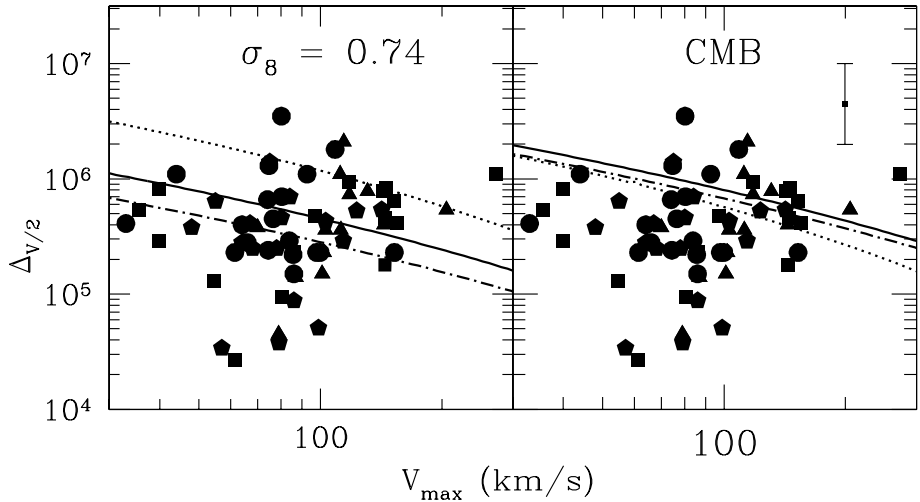}
\caption{ 
The central density parameter as a function of the maximum halo
circular velocity.  Dotted, solid, and dash-dot lines in each panel
refer to $\wQ = -0.5$,$-1$, and $-1.5$, respectively.  The left panel
is for a fixed $\sigma_8=0.74$ and the right panel is normalize to
match the CMB: $\sigma_8 \simeq 0.6$ , $0.9$, and $1.0$ for $\wQ =
-0.5$,$-1$, and $-1.5$, respectively. The error bar in the right panel
shows the expected theoretical $1-\sigma$ scatter in $\Delta_{V/2}$
due to the scatter in $\cvir$. The points are for observed LSB and
dwarf galaxies (see text for references).
}
\label{fig:DeltaVby2}
\end{figure*}

We compare the B01 model predictions in terms of these quantities in
Figure~\ref{fig:DeltaVby2} for three representative quintessence
cosmologies to the observational data of low-surface brightness and
dwarf galaxies compiled by Zentner \& Bullock (2002; the same set of
observed galaxies are discussed in Hayashi et al. 2003, who reach
similar conclusions) from the observational work of Swaters (1999), de
Blok, McGaugh, \& Rubin (2001), and de Blok and Bosma (2002). The
error bar in the right panel shows the theoretical $1\sigma$ scatter
in $\Delta_{V/2}$ expected at fixed $V_{\rm max}$ due to the scatter
in $\cvir$.

While it is difficult to quantify the agreement of each quintessence
cosmology with observational data, the cosmological model that is most
commonly referred to as the standard, concordance cosmological model
with $\wQ = -1$ $\sigma_8 = 0.9$ (solid line, right panel) seems to be
in conflict with the extant observational data.  For $w=-1$, lowering
the normalization of the power spectrum to the value $\sigma_8 \approx
0.74$ (solid line, left panel) can greatly alleviate this discrepancy
(Alam, Bullock, \& Weinberg 2002; Zentner \& Bullock 2002; McGaugh
\etal 2003).  However, a $w=-0.5$ model with the same $\sigma_8
\approx 0.74$ normalization (dotted line, left panel) does not do as
well because earlier structure formation produces higher galactic
central densities.

Notice that the trends with $\Delta_{V/2}$ for fixed $V_{\rm max}$ do
not scale as might be expected from the $\cvir$ trends at fixed mass
shown in Figure~\ref{fig:cvirplot}.  This is because $\Delta_{V/2}$ is
a physical measure of density and it increases not only with $\cvir$
but also with $\DeltaVir$ (i.e. haloes are defined with respect to
different overdensities).  This effect is most apparent when comparing
the right panels of Figures~\ref{fig:cvirplot} and
~\ref{fig:DeltaVby2}.  Though the concentrations of haloes with
$w=-0.5$ and $\sigma_8=0.6$ (dotted-line, right panel, Figure
~\ref{fig:cvirplot}) are much lower than those in the standard
$w=-1.0$, $\sigma_8 = 0.9$ case (solid line, right panel), the actual
densities of those haloes are roughly the same in the right panel of
Figure~\ref{fig:DeltaVby2}.  This is because $w=-0.5$ models have
higher virial densities (see Figure~4).  Even though haloes in the
low-normalization $w=-0.5$ model tend to have the same
(rapid-collapse) formation epoch as those in the higher-normalization
$w=-1$ model, the higher virial densities in the former model make the
haloes denser overall.

By inspecting Figure ~\ref{fig:DeltaVby2}, we can immediately
determine that models with $\wQ < -1$ and moderately low $\sigma_8$
($\sigma_8 \sim 0.7-0.8$) can bring theoretical predictions to rough
agreement with rotation curve data from low surface brightness and
dwarf galaxies.  Though it is clear that sufficiently decreasing
$\sigma_8$ can bring any model into accord with the median of the
data, a $w = -0.5$ model would require $\sigma_8 < 0.6$.  Such a low
normalization would be nearly impossible to reconcile with $z=0$
cluster abundance data.  Thus from the standpoint of quintessence,
models with $\wQ < -1$ seem mildly favored by galaxy density data.
Conversely, models with $\wQ$ as high as $\sim -0.5$ are strongly
disfavored by galactic rotation curves coupled with only a weak prior
on the normalization of the power spectrum.

Note that none of the models can easily account for the very low data
points.  Nevertheless, the scatter in the data is not extremely large
compared to the scatter expected from the halo-to-halo variations
observed in N-body simulations.  For example, at $V_{\rm max} = 80$
km/s, the $1\sigma$ scatter in N-body simulations is
$\sigma(\log(\Delta_{V/2})) \simeq 0.37$ while the $1\sigma$ scatter
in {\em all} $67$ data points is $\sigma(\log(\Delta_{V/2})) \simeq
0.41$.  This suggests that lowering the median of the theoretically
predicted central densities, perhaps by a reduction in $\sigma_8$ or
invoking a tilted or running power spectrum that reduces power on
galaxy scales, or as we discuss here, by invoking $\wQ < -1$
quintessence, may be sufficient to bring the predictions into good
agreement with the data.  Yet, we must bear in mind that our
calculations are approximate. The most obvious omission is that all of
our calculations are based on N-body simulations that contain no
baryons.  The effects of baryonic contraction are likely to be small
in LSB galaxies (\eg de Blok \& McGaugh 1997) and would tend to drive
rotation curves to higher values or $\Delta_{V/2}$ and $V_{\rm max}$
in the simplest models (Blumenthal et al. 1986).  This serves only to
increase the apparent discrepancy.  Additionally, rotation curve
measurements may yet be subject to poorly-understood systematic
effects in the reduction of the observational data (Swaters \etal
2003).  Currently, it is difficult to draw a firm conclusion.

\section{Summary and Conclusions}
\label{sec:conclusions}

Although the nature of dark energy is unknown, its effects on
structure formation can be studied using numerical N-body
simulations. We have performed a series of these simulations for a
range of dark energy equation of state parameters. Confirming previous
findings by Linder \& Jenkins (2003), Klypin \etal (2003), Macci\`o
\etal (2003), and \L okas \etal (2003) we show that the J01 formula
provides a good fit to halo mass functions even in the presence of
non-\lcdm dark energy. We show that this is true for models with
$\wQ<-1$ as well.

The density structure of dark matter haloes is also affected by dark
energy. We have shown how the predictions of the B01 model are
modified when dark energy with constant $\wQ$ is accounted for.  As
structure tends to form earlier in models with less negative $\wQ$,
halo concentrations tend to be somewhat higher in these models. These
findings are in agreement with the results of Klypin \etal (2003) and
qualitatively agree with Dolag \etal (2003), although we probe a
different range of masses than the latter. The larger number of haloes
with NFW profile fits and concentrations in our study allows us to
quantitatively test the B01 model. We find that the original ($F=0.01,
K=4.0$) over-predicts the concentrations of haloes in our simulations
by about $\sim 12-15\%$. However, the shape of the mass-concentration
relation that we find is the same as in B01, and we find that a
slightly modified set of the B01 parameters ($F=0.01, K=3.5$) matches
our haloes well. This offset may likely be caused by the lower force
resolution of our GADGET simulations compared to the
adaptive-refinement code ART used in B01.  For the haloes in our
simulations the adopted B01 model accurately reproduces the median
concentration-mass relation over a range of masses from $M_{\rm vir}
\sim 6 \times 10^{11}$ to $M_{\rm vir} \sim 4 \times 10^{13}\hinv
\msun$.  We confirm that for a fixed mass halo concentration decrease
with redshift as $1/(1+z)$, at least out to $z \sim 2.5$.

Interestingly, we find that halo concentrations are more easily
understood when the halo virial radius is defined in terms of a
cosmology-dependent virial overdensity rather than by one that uses a
fixed overdensity of $\rho/\rho_{\rm crit}=200$.  The result supports
one of the (previously-untested) assumptions of the original B01
model.  Specifically, we argue that halo densities in different
cosmological models are influenced both by changes in the overall
virialization process of haloes as well as by changes in epoch when
the halo cores collapse.  As noted in \S\ref{sec:DeltaVby2}, it is
important to include both of these physical processes when comparing
predictions for galaxy densities to real data, as in
Figure~\ref{fig:DeltaVby2}.

Having confirmed that the B01 model correctly describes the scaling of
halo concentrations as a function of mass and redshift even in
cosmologies with $w\ne-1$, we have investigated the effects of dark
energy on a comparison between model predictions and observations of
central halo densities. Zentner and Bullock (2002) found that the
observed distribution of $\Delta_{\rm V/2}$ as a function of $V_{\rm
max}$ is inconsistent with the predictions of the B01 model for
$\Lambda$CDM and $\sigma_8=0.9$. The model predicts haloes that are
simply too concentrated (see also Primack 2003). A lower value of
$\sigma_8=0.75$, as used in our simulations, can alleviate this
discrepancy, but such models may face other difficulties regarding
early reionization (Somerville, Bullock, \& Livio 2003) and possibly
with reproducing the properties of halo substructure (Zentner \&
Bullock 2003).  Including the effects of dark energy, we find that for
models with $w>-1$ the problem is exacerbated because haloes collapse
earlier and because they have higher virial overdensities.  Note that
for the extreme case of $w=-0.5$, even a normalization as low as
$\sigma_8 = 0.6$ seems disfavored by the data. Thus one interesting
conclusion is that the rotation curves of galaxies coupled only with a
weak prior on the normalization of the power spectrum of density
fluctuations seem to disfavor quintessence models with $\wQ$
significantly larger than $-1$ without measuring the expansion history
of the Universe, as is done in SNIa analyses. Models with $w<-1$ do
better, and can tolerate higher values of $\sigma_8 \sim 0.8$, but not
high as high as $\sigma_8 \sim 1$, as is suggested the CMB
normalization.~\footnote{Note that using $K=3.5$ instead of $K=4.0$
would not change any of the conclusions regarding galaxy rotation
curves and the central density problem. }

As we have pointed out throughout this paper, there have been a number
of previous studies of the effects of dark energy ($\wQ \ne -1$) on
dark matter halo abundances and concentrations. Here we summarize how
this work distinguishes itself from these past efforts. We have shown
for the first time how results for $\Delta_{\rm vir}$ and the linear
transfer function are modified for dark energy with $w<-1$. In
addition we have expanded on previous studies (Schuecker et al. 2003)
of $\delta_c$ in phantom cosmologies. We have pointed out that there
are significant differences in the amplitude and $\wQ$-dependence of
$\sigma_8$ between normalizations based on the CMB and galaxy cluster
abundances. For the first time we have been able to draw a distinction
between a definition of concentration based on a cosmology-dependent
$R_{\rm vir}$ and constant $R_{200}$, and found that our simulations
prefer the former. Unlike previous studies of the effects of dark
energy on halo concentrations (Klypin et al. 2003; Dolag et al. 2003),
which analysed a small number of pre-selected haloes, we have
performed an analysis of $\sim1600$ haloes in each simulation. This
allowed us to directly test the B01 model for $c(M,z)$ with
statistical significance. Furthermore these previous studies focused
on group- and cluster-sized haloes, whereas we have extended this
study down to galaxy mass haloes. Lastly, we have shown for the first
time how different values of $\wQ$ and power spectrum normalizations
affect the theoretical predictions of $\Delta_{V/2}$ vs. $V_{\rm
max}$. A comparison to observations shows that models with w as low as
-0.50 are strongly disfavored for any value of $\sigma_8$, standard
$\Lambda$CDM models require $\sigma_8 \la 0.8$, and models with
$\wQ<-1$ can accomodate higher values of $\sigma_8$.

As future observations further constrain the nature of dark energy, it
will become necessary to extend these types of studies to more
realistic models of dark energy.  Future simulations with higher mass
and force resolution, including the effects of baryons, and
incorporating more realistic dark energy models will further advance
our understanding of the interplay between cosmology and dark matter
halo structure.

\medskip

We thank P.~Madau for interesting and helpful discussions.  MK is
supported by NSF grant AST-0205738. LES is supported by the Department
of Energy grant DE-FG02-91ER40690.  ARZ is supported by The Center for
Cosmological Physics at The University of Chicago under NSF PHY
0114422.  JSB is supported by NASA through Hubble Fellowship grant
HF-01146.01-A from the Space Telescope Science Institute, which is
operated by the Association of Universities for Research in Astronomy,
Incorporated, under NASA contract NAS5-26555.  JRP is supported by NSF
grant AST-0205944. ARZ thanks The Center for Cosmology and Particle
Physics at New York University for their hospitality during several
visits while this work was in progress.  We thank V. Springel for use
of the publicly-available code GADGET and we than U. Seljak and
M. Zaldarriaga for use of the publicly-available code CMBFAST.

\label{lastpage}

\end{document}